\documentclass[twocolumn,twoside]{revtex4}
\usepackage{graphicx}
\usepackage{color}
\usepackage{fancyhdr}
\usepackage{amsmath}
\usepackage{tikz}
\usetikzlibrary{decorations.pathmorphing}
\usetikzlibrary{decorations.markings}
\begin{document}

\title{Lepton (non-) unversality in (flavor changing) neutral current B decays}

%

\author{\bf Rodrigo Alonso}
\affiliation{Kavli Institute for the Physics and Mathematics of the Universe (WPI), The University of Tokyo Institutes for Advanced Study, The University of Tokyo, Kashiwa, Chiba 277-8583, Japan}

\begin{abstract}
	{\color{blue}\bf May 06-10, 2019}
The following proceedings contain a theory perspective on the flavor changing neutral current anomalies reported by LHCb on the ratios $B\to K^{(*)} \mu\mu/B\to K^{(*)} ee$.
\end{abstract}

\maketitle

\thispagestyle{fancy}


\section{Introduction}
Next year will mark the 50th anniversary of the seminal paper by Glashow, Iliopoulos and Maiani~\cite{Glashow:1970gm} and, while much experimental progress has been made in the ellapsed time, our fundamental theory understanding of flavor has not changed since. This speaks to the insight of the aforementioned work but is not to say there has been no effort from the theory community; new ideas have emerged as answers for a deeper understanding, from spontaneous symmetry breaking to extra-dimensions. At the same time the wealth of data from flavor experiments provided the foundation for new fields of study like heavy quark effective field theory or soft collinear effective field theory while yielding valuable input on theories beyond the established one. 

In this regard the experimental program on B physics has been at the forefront of the progress in the field of flavor with sustained relevance throughout the last decades and into the future with experiments like LHCb and Belle II. It might be also the source of momentous experimental results if the present deviations from the standard theory are confirmed opening the door for the long overdue breakthrough in flavor physics.

\section{Theory context}

Flavor as it is understood today originates from the (different) mass of same-charge elementary particles. The spectrum of fundamental particles presents the theoretically self-consistent structure of five different-charge particles replicated three times. Each of these three structures consists of three color-charged fermions (quarks) and two color neutral fermions (leptons), all five of them charged under the electroweak group. Explicitly
\begin{gather} \nonumber
q_L\sim (3,2,1/6) \quad
u_R \sim (3,1,2/3) \quad
d_R \sim (3,1,-1/3)\\
\ell_L \sim (1,2,-1/2)\qquad e_R \sim (1,1,-1)\label{SMReps}
\end{gather}
where charge is encoded as representation under the color ($SU(3)_c$), weak isospin ($SU(2)_L$) or hyper-charge ($U(1)_Y$) interactions as (Rep. under $SU(3)_c$, Rep. under $SU(2)_L$, Rep. under $U(1)_Y$) and the sub-index $L,R$ indicates the chirality (helicity at high energy) i.e. Lorentz group $SO(4)\sim SU(2)_L\times SU(2)_R$ representation. 

Interactions are given by particle's charges (representations) and controlled by the respective magnitude $g_c,g, g_Y$ of strong, weak-isospin and hypercharge couplings. While representations under a non-abelian group are discreet, charges under the Abelian $U(1)_Y$ constitute  a priori a continuum. Nevertheless for the quantum consistency of the theory anomaly cancellation is required which imposes, given the non-abelian representations, the ratios between the different hyper-charges given in~(\ref{SMReps}).
These are however the similarities between the three copies of generations, to account for flavor one has to explain the differences.

\begin{figure}[h]
\includegraphics[width=.45\textwidth]{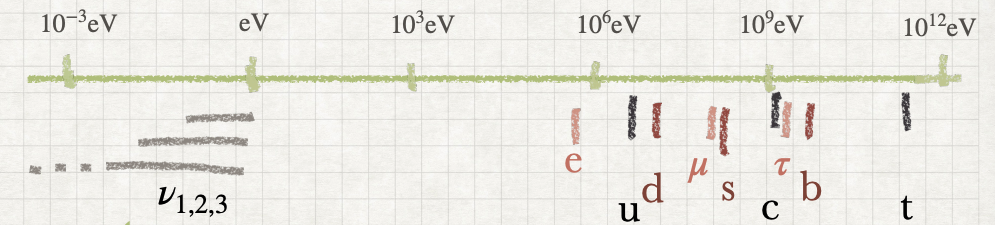}
\caption{Sketch of the fermionic spectrum}\label{Sktch}
\end{figure}

As outlined in the introduction, the difference is mass, in a sufficiently broad acceptation. The conflict within the Standard Model(SM) is that mass cannot be generated for the `theory' outlined above. Mass conflicts with charge conservation of the electroweak symmetry group, yet is an observed property of the low energy world which we experience. The reconciliation between the two is brought about by the Higgs doublet, $H\sim (1,2,1/2)$, and spontaneous symmetry breaking, i.e. the `concealment' of the electroweak group at low energies. The Higgs scalar field allows for an interaction between different-charge fermions that respects the electroweak symmetry as:
\begin{align}\nonumber
S_Y=&-\int d^4x\left\{\bar q_L Y_u \tilde H u_R +\bar q_L Y_d Hd_R +\bar\ell_L Y_e  He_R\right\}\\
& +{\rm h.c.}\label{Eq:Yuk}
\end{align}
while the vacuum state of the theory has a non-vanishing background value for the Higgs field $\left\langle H\right\rangle=(0,v/\sqrt2)$, $v=246$GeV. This makes the electroweak symmetry not manifest at low energies and yields masses for three of the 4 force carriers of $SU(2)_L\times U(1)_Y$, $W^{\pm}$ and $Z$ leaving only the photon intact. At the same time eq.~(\ref{Eq:Yuk}), when expanded around the vacuum yields a mass term for fermions:
\begin{align}
M_u=&\frac{v}{\sqrt2}Y_u=V^\dagger{\rm diag}(m_u)\,,\\
M_d=&\frac{v}{\sqrt2}Y_d={\rm diag}(m_d)\,,\\
M_e=&\frac{v}{\sqrt2}Y_e={\rm diag}(m_e)\,,
\end{align}
where each entry is a $3\times 3$ matrix for the 3 generations of matter and we have, in the second equality, used the freedom in rotating $q_L,u_R,d_r,\ell,_L,e_R$ fields in flavor 3-space to reduce these matrices to the physical parameters. The diagonal matrices encode the masses for the different generations and present a strong hierarchical pattern for all (electromagnetically) charged fermions as shown in the fig.~\ref{Sktch}.
V is the Cabibbo-Kobayashi-Maskawa unitary mixing matrix with 3 angles and a phase, and the equivalent matrix for neutrinos is
the Pontecorvo-Maki-Nakagawa-Sakata mixing matrix with another 3 angles and a variable number of phases depending on the nature of neutrino masses. The quark mixing angles are small and $V$ a nearly identity-like matrix while opposite stand leptons with large mixing angles and yet to be determine phases. Combine this with neutrinos lying 6 orders of magnitude in mass below the lightest charged fermion and one has a pattern of masses and mixings which is intricate, seemingly capricious and theoretically unaccounted for.

In the vacuum state we can tell the two components of the $SU(2)_L$ doublets $q_L=(u_L,d_L)$  apart and in particular rotate them in flavour space separately to diagonalize mass matrices and shift the unitary matrices to $W$ boson couplings as
\begin{align}
\frac{g}{\sqrt2}\, \bar u_L V \gamma^\mu d_L W_{\mu}^++{\rm h.c.}
\end{align}
while neutral bosons $Z$ and photon couple universally to fermions and the current $\Sigma_i\bar u_{L_i}\gamma u_{L_i}$ stays unchanged under rotations. This fact, which might seem a passing remark, has very relevant phenomenological implications. Different generations change into each other through their couplings to the W boson,  and for example the decay of a third-generation-flavored particle like the B meson will predominantly occur through a `charged current'.

The case of a decay via a neutral current is still possible at the quantum level via virtual charged currents as
\begin{align}
\raisebox{-9.5mm}{\begin{tikzpicture}
	\draw [thick]  (-1,1) node [anchor=north east] {$d_i$}--(0,0);
	\draw [thick] (-1,-1) node [anchor= south east] { $d_j$} -- (0,0);
	\draw [style={decorate, decoration={snake}},thick] (0,0) 
	-- (.75,0);
		\draw [style={decorate, decoration={snake}},thick] (-0.6,0.6) node [anchor=west] {$\,\,gV$} -- (-0.6,-0.6)  node [anchor=west] {$gV^\dagger$};
	\end{tikzpicture}}\qquad \sim\,\,
g\frac{g^2}{(4\pi)^2}\frac{V^\dagger_{i k} m^2_{u_k}V_{kj}}{M_W^2} \,,
\end{align}
where the case of down-type external states has a suppression due to small off-diagonal elements in $V$, the case of external up-type has small ratios of masses $m_d/M_W$ as an extra suppression and finally charged leptons in the external states have a very acute small mass suppression $m_\nu/M_W$. 

The couplings we have discussed so far, which basically constitute the SM, are however not expected to be the sole interactions that matter possesses. Rather we think of them as the leading terms in a low energy expansion on energy, $E$, with order parameter $E/\Lambda$ and $\Lambda$ a new scale $\geq 246$GeV. In this way a correction to the $Z$ coupling to e.g. leptons reads
\begin{align}
\frac{C}{\Lambda^2}\bar \ell_L \gamma_\mu \ell_L H^\dagger i D H +{\rm h.c.}
\end{align}
which modifies the $Z$ boson interactions as $g(1+C v^2/\Lambda^2)$ and with $C$ a $3\times 3$ matrix in flavor space. 
The flavor of $Cv^2/\Lambda^2$ is however highly constrained experimentally; LEP data constrains non-universality ($C_{ii}\neq C_{jj}$) at the per mile level whereas flavor changing $C_{i<j}$ at the per million level.

At lower energies one can test the flavor of these same neutral currents with flavored hadron decays, which in the Standard Model are mediated by EW bosons as, for example,
\begin{align}
\raisebox{-9.5mm}{\begin{tikzpicture}
	\draw [thick]  (-.75,1) node [anchor=north east] {$d_i$}--(0,0);
	\draw [thick] (-.75,-1) node [anchor= south east] { $d_j$} -- (0,0);
	\draw [style={decorate, decoration={snake}},thick] (0,0) 
	-- (.75,0);
	\draw [style={decorate, decoration={snake}},thick] (-0.6,0.75) node [anchor=west] {$\,\,gV$} -- (-0.6,-0.75)  node [anchor=west] {$gV^\dagger$};
	\draw [thick] (.75,0) --(1.25,1) node [anchor =east] {$e$};
		\draw [thick] (.75,0) --(1.25,-1) node [anchor=east] {$e$};
	\end{tikzpicture}}\quad \sim\,\,
\frac{g^2}{(4\pi)^2}\frac{V^\dagger_{i k} m^2_{u_k}V_{kj}}{M_W^2}G_F
\,,\label{SM4F}
\end{align}
which would generate a four fermion interactions operator like $\bar d_{L_i} \gamma^\mu d_{L_j}  \bar e\gamma^\mu e$.
To these again one expects corrections from physics beyond the SM which we will study in the next section.


\section{Experimental evidence}
In the study of hadrons,
connecting the elements of theory with experiment requires the further step of taking matrix elements. This step is no hurdle in a perturbative theory but the strong character of color interactions at low energies makes it impossible to compute the result from first principles. One then parametrizes its ignorance in form-factors which can be estimated in certain limits or in some cases given by the lattice but eventually it is their uncertainty that rules the theory prediction. A way to avoid these uncertainties is to,  in some sense, use flavor to `factor them out'; consider leptonic or semileptonic decays of $B$ mesons to light leptons ($\mu,e$), we have that in this case the number of form factors collapses into one:
\begin{align}\nonumber
\left\langle  K(k) \right| \bar s\gamma^\mu b_L \left|B(p)\right\rangle &=f_+(q^2)(p+k)^\mu+\mathcal O(m_\ell/m_B)\\
\left\langle 0\right| \bar s\gamma^\mu b_L \left| B_s(p)\right\rangle &=f_{B_s}p^\mu+\mathcal O(m_\ell/m_B)
\end{align}
where by $\mathcal O( m_\ell/m_B)$ we refer to terms which, when contracted with the leptonic matrix element, will be of such order compared to the leading term.
One can now take the ratio of two different final lepton decays to cancel out the form factors and in the process obtain a probe of lepton universality in $B$ decays. This probe is more straightforward in the semi-leptonic case given that the leptonic decay scales with lepton mass due to angular momentum conservation.

One therefore has a clean probe in the search for new physics given that this ratio in the SM is generated by diagrams as the one in ~(\ref{SM4F}) and equal 1 at leading order.
Experimental data on two such ratios from LHCb does indeed deviate at present from 1~\cite{Aaij:2019wad}:
\begin{align} \nonumber
R_K\equiv\frac{B\to K \mu\mu}{B\to K ee}=0.846\,\,\substack{+0.060 \\ -0.054} ({\rm stat}) \substack{+0.014 \\ -0.016} ({\rm sys})
\end{align}
and~\cite{Aaij:2017vbb}
\begin{align} \nonumber
R_K\equiv\frac{B\to K^* \mu\mu}{B\to K^* ee}=0.69\,\,\substack{+0.11 \\ -0.07} ({\rm stat}) \substack{+0.05 \\ -0.05} ({\rm sys})
\end{align}
both in the $1.1$GeV$<q^2<6$GeV bin (LHCb also reports a deviation in a low-$q^2$ bin for $R_K^*$~\cite{Aaij:2017vbb}).
The theory prediction in the SM has negligible uncertainty~\cite{Bordone:2016gaq} and each of the ratios presents a $~2.5\sigma$ significance for a deviation from the SM.
It should also be remarked that a value for $R_K^*$ has been reported by Belle~\cite{Abdesselam:2019wac} although its uncertainty at present makes it compatible with both the SM and LHCb.

There are in addition significant deviations in angular observables in the muon channel with however higher theory uncertainties. All in all the data paints a picture for a new effect which deserves scrutiny.

Before plunging into an analysis however let us inspect the type of physics that the new effect displays:
\begin{itemize}
\item Flavor-changing in the quark sector
\item Flavor-conserving but universality-violating in the lepton sector
\item Magnitude around to the Standard Model
\end{itemize}
Flavor changing effects are indeed, for the reasons elaborated in the previous section, a powerful probe into new physics given the low SM model `background' but then why does the effect has the size of the SM contribution and is not one, two orders of magnitude above? This could be because it is produced by particles of masses not far from the EW scale and connected to the hierarchy problem. On the other hand, all points above can be summarized in, why have we not seen this effect elsewhere, e.g. purely hadronic or leptonic physics? We do not have the answers to these questions, and that is what brightens the prospects in the field.

\section{Analysis}
The model-independent analysis of new physics behind the $R_{K^{(*)}}$ anomalies is analogous to the discussion in sec.~II, let us sketch it here while a more in depth review can be found in e.g.~\cite{Cerri:2018ypt}.
At the $B$-meson scale the electroweak-boson mediated processes that induce flavor changing neutral current decays (e.g.~\ref{SM4F}) take the form of 4 fermion contact interactions given that the $W,Z$ are not kinematically accesible.
This is to say that the relevant interactions read:
\begin{equation}
\label{eq:Leffnc}
S_{\rm n.c.}=
\int d^4x \frac{4 G_F}{\sqrt{2}}V^\dagger _{bt}V_{ts} C_k  \mathcal{O}_k,
\end{equation}
where the basis contains 4 fermion operators generated by the SM like
\begin{align}\label{eq:semilOps}
\mathcal{O}_9  &= 
\frac{e^2}{(4 \pi)^2} [\bar s \gamma_\mu  P_ Lb][\bar{l} \gamma^\mu l],\\
\mathcal{O}_{10}  = &
\frac{e^2}{(4 \pi)^2} [\bar s \gamma_\mu  P_L b][\bar{l} \gamma^\mu \gamma_5 l], 
\end{align}
with  $C_9^{SM}=4.24\simeq -C_{10}^{SM} $ but in generality one considers {\bf all} other possible four fermion structures including deviations in the above, their chiral partners $\mathcal{O}_{9,10}^\prime=\mathcal{O}_{9,10}(L\to R)$ and~\cite{Bobeth:2007dw}:
\begin{align}\label{eq:nonSM:scalars} 
\mathcal{O}_S^{(\prime)} & = \frac{e^2}{(4 \pi)^2} [\bar s P_{R(L)} b] [\bar{l} l], \\
\mathcal{O}_P^{(\prime)}  &= \frac{e^2}{(4 \pi)^2} [\bar s P_{R(L)} b] [\bar{l} \gamma_5 l],
\\  
\mathcal{O}_T   & = \frac{e^2}{(4 \pi)^2} [\bar s \sigma_{\mu\nu} b][\bar{l} \sigma^{\mu\nu} l], \\ \label{eq:nonSM:tensors}
\mathcal{O}_{T5} & = \frac{e^2}{(4 \pi)^2} [\bar s \sigma_{\mu\nu} b][\bar{l} \sigma^{\mu\nu} \gamma_5 l],
\end{align}
where we note a loop factor $\alpha_{\rm em}/4\pi=e^2/(4\pi)^2$ is used in the definition of the operators and hence we expect  a  coefficient of $16\pi^2 v^2/\Lambda^2$ from tree level new physics of mass $\Lambda$. This points at masses of $\Lambda\sim30$TeV given the size of deviations.

A closer look narrows down the possible operators, at least for a sizable improvement of the new physics statistical agreement with data over the SM.
One first thing to notice is that if the scale of new physics is above the EW scale it must admit a description in terms of representations of the full SM group $SU(3)_c\times SU(2)_L\times U(1)_Y$. So we can try and promote the fermion fields in  the operators of (\ref{eq:semilOps}-\ref{eq:nonSM:tensors}) to full representations, take a tensor operator for example
\begin{align}
\underset{U(1)_Y:}{} \,\,\,\,\underset{1/3}{\bar d_R}\sigma_{\mu\nu} \underset{1/6}{\bar q_L} \underset{1}{\bar e_R} \sigma_{\mu\nu} \underset{-1/2}{ \ell_L}  \,\,\,\,\underset{\Sigma \neq 0}{}
\end{align}
so the hyper-charge does not add up to one and given that $\sigma_{\mu\nu}P_L\otimes \sigma_{\mu\nu}P_R=0$ we find there is no operator producing such a structure at dimension 6.
It can be generated at dimension 8 with the insertion of Higgs doublets but given that it is a sub-sub-leading effect we discard it here. The rest of operators can be generated from the fully gauge invariant basis:
\begin{align}
Q_{\ell q}^{(1)}          =      & (\overline q \gamma^\mu q_L)(\bar \ell \gamma_\mu \ell_L) &
Q_{\ell q}^{(3)}          =      & (\overline q\vec{\tau} \gamma^\mu q_L)\cdot(\bar \ell \vec{\tau} \gamma_\mu \ell_L)\nonumber\\
Q_{\ell d}        =       & (\bar d \gamma^\mu d_R)(\bar \ell \gamma_\mu \ell_L)&
Q_{qe}     =          & (\overline q \gamma_\mu q_L)(\overline e \gamma^\mu e_R)   \nonumber\\
Q_{ed}   =                   & (\bar d_R \gamma^\mu d_R)(\overline e \gamma_\mu e_R) &
Q_{\ell edq} =& (\bar \ell_L e_R)(\overline d_R q) \label{eq:Leffdown}
\end{align}
yet we note there are two scalar operators above only ($Q_{\ell edq}$ and its h.c.) vs 4 in eqs.(14-15) so their coefficient are related by gauge invariance as:
\begin{align}
C_P&=-C_S\,, & C_P^\prime&=-C_S^\prime\,.
\end{align}
\begin{figure}[h]
	\includegraphics[width=.45\textwidth]{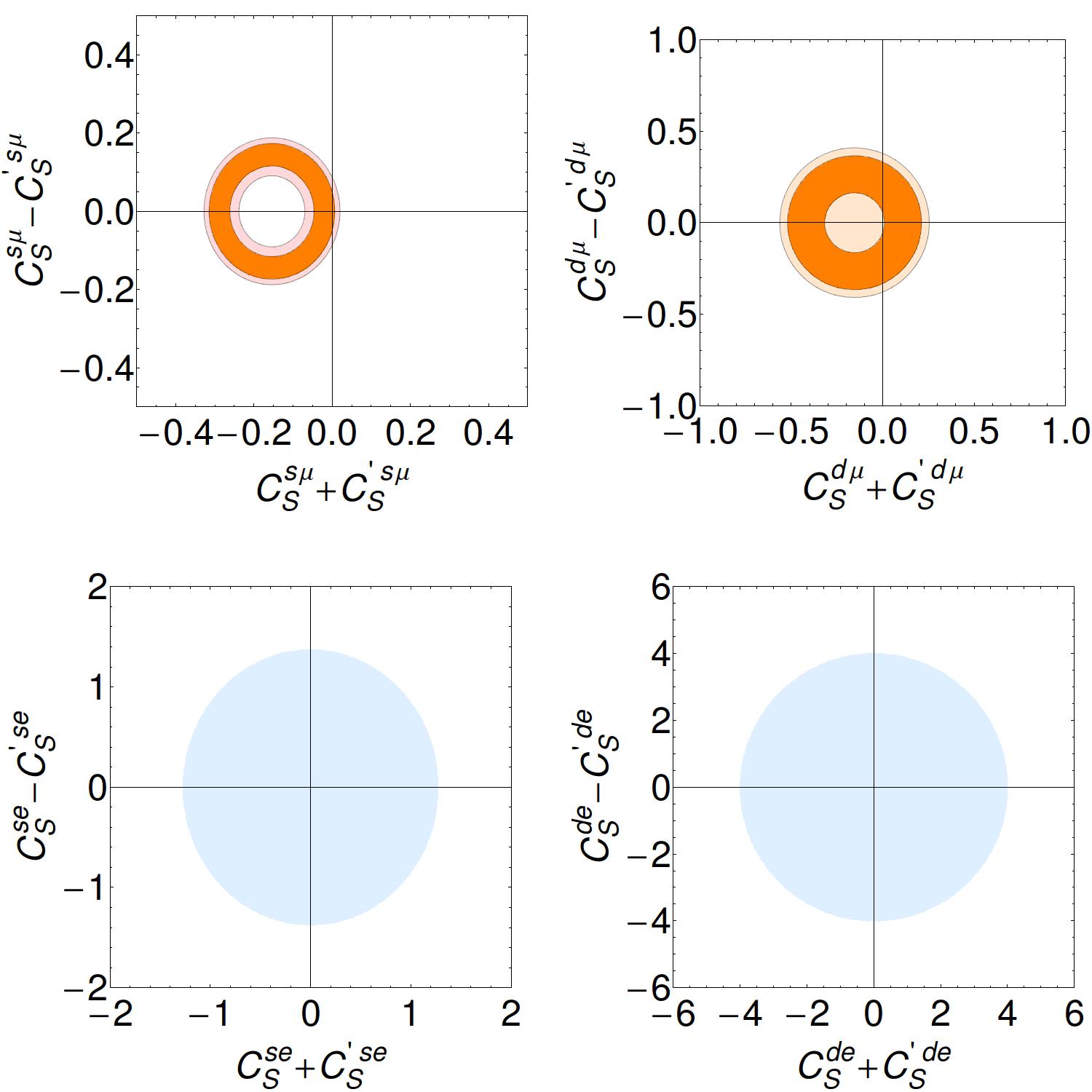}
	\caption{Bounds on scalar ops from leptonic $B_{s,d}$ decays}\label{CSP}
\end{figure}
Furthermore the two remaining scalar paremeters $C_S, C_S^\prime$ contribute to fully leptonic decays without the SM chiral suppression and hence subject to severe bounds as displayed in fig.~\ref{CSP}. 
This is so much so~\cite{Alonso:2014csa} that scalar operators cannot acount for the deviations in $R_{K,K^*}$.

One is therefore left with current-current operators only. Here consideration of the fact that there is deficit in both $K$ and $K^*$ modes which are excited by the vector and axial currents respectively, implies that new physics shares approximately the chiral structure of the SM. This selects $\mathcal O_{9,10}$ and discards $\mathcal O_{9,10}^{\prime}$, see e.g.~\cite{DAmico:2017mtc,Geng:2017svp}. Finally the latest data has preference for a chiral current in the lepton sector too, $\delta C_9=-\delta C_{10}$~\cite{Aebischer:2019mlg} leaving us at the end of this analysis with one operator (parameter) only.

The question of why does the effect has the SM magnitude therefore becomes more acute; it has not only the SM size but also its chirality structure. Perhaps this is indicative of all sources of flavor in nature residing in left-handed currents?
The more pressing question is nonetheless related to the other points raised in sec.~III as to why has this effect not appeared in other processes and whether it is even compatible with other bounds. 
One should indeed inspect other very precisely measured processes in e.g. kaon physics, lepton universality violating $Z$ decays or lepton flavor violating decays. The issue with these other processes is that their modification is model dependent, yet one can stick to the effective analysis and minimal case of operators $\mathcal O_{9,10}$ only and estimate their loop level phenomenology via renormalization group evolution~\cite{Buras:2014fpa,Feruglio:2017rjo}. The result is that the $R_{K^{(*)}}$ anomalies can be accommodated while while compatible with other constraints.

The next step in the analysis is to inspect the possible UV completions that would produce the 4 fermion operators and the searches for new resonances. `Opening up' the operator as in the classification of the seesaw model types, yields two possibilites, $Z^\prime$ models or leptoquarks. Their study however falls beyond the present letter, see {\bf WedE1400} in this conference's proceedings.

Before closing however we would like to point out a possible explanation for the leptonic structure of the anomalies and its testing. It has been argued that lepton universality violation brings about lepton flavor violation~\cite{Glashow:2014iga}. While neutrinos massiveness is certainly a source of lepton flavor violation in practice new physics might have a flavor structure disconnected from it. If so one can motivate with symmetry the absence of flavor violation while allowing for non-universality by postulating that the new physics respects a $U(1)_e\times U(1)_\mu\times U(1)_\tau$ symmetry, that is an independent phase rotation for each generation. This symmetry is defined in the mass basis of the charged leptons so one could argue that this proposal is simply the alignment of new physics flavor with the mass of charged leptons. However, this alignment is not necessary if {\it the source of flavor in the new phyis is the charged lepton mass matrix}, that is $Y_e$ in eq.~\ref{Eq:Yuk}. This idea is indeed realized in minimal flavor violation and can be formulated in terms of a bigger encompassing flavor group. The reader is referred to~\cite{Alonso:2015sja} for the fully fleshed formulation, here we shortcut to the consequences in the coefficients for operators $\mathcal O_{9,10}$
\begin{align}
\frac{e^2}{(4\pi)^2}V^\dagger _{bt}V_{ts}\left( \delta C_9\right)_\alpha=-\frac{v^2}{\Lambda^2}\frac{m_\alpha^2}{m_\tau^2} \label{eqMLFV}
\end{align}
with $\alpha$ marking lepton flavor $\alpha=e,\mu,\tau$  and $\delta C_{10}=-\delta C_9$. 
\begin{figure}[h]
	\centering
	\includegraphics[width=80mm]{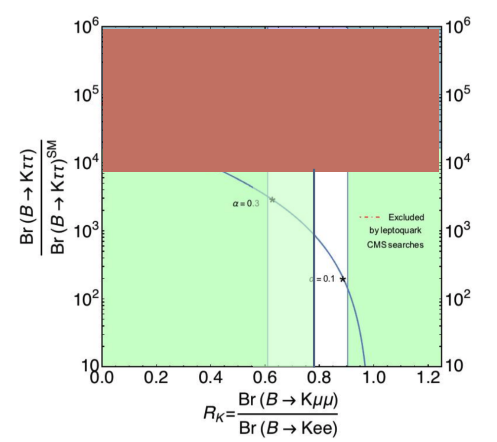}
	\caption{Correlation between different lepton flavor observables in the scenario of eq.~(\ref{eqMLFV}).} \label{MLFVFig}
\end{figure}
In this way the anomalies in $R_{K^{(*)}}$ can be accommodated by a modification of the muon channel yet an immediate consequence is an $O(10^3)$ enhancement in tauonic processes. Surprising as it might seem this is compatible with present data, see fig.~\ref{MLFVFig} and furthermore has the right size to accommodate the deivations in ratios in the charged current decays $B\to D^{(*)}\ell \nu$,~\cite{Alonso:2015sja}.

\section{Summary}
The present letter was meant as a brief and colloquial presentation of the theory perspective on the flavor changing neutral current anomalies $R_{K^{(*)}}$. The emphasis was on the questions that these observations pose for the theorist and an example of how a possible answer can be tested experimentally. 
\begin{acknowledgments}
This work was supported by World Premier International Research Center Initiative (WPI Initiative), MEXT, Japan.
\end{acknowledgments}

\bigskip 

\end{document}